\magnification\magstep0
\hsize = 12.0 cm
\vsize = 18.6 cm
\overfullrule 0pt
%\hoffset=0.70 true cm
%\voffset=1.10 true cm
%
%\magnification\magstep2
%\hsize= 16 true cm
%\vsize= 24 true cm
%\vsize= 22.0 true cm
%\hoffset=-0.70 true cm
%\voffset=-1.10 true cm
%
\nopagenumbers
%\headline={\hss\tenrm\folio}
\tenrm

\noindent{\bf Projected Dynamics for Metastable Decay in Ising Models }
\vskip  0.15 true cm
\noindent{M.~Kolesik$^{1,2}$, 
          M.~A.~Novotny$^{1,3}$, 
          P.~A.~Rikvold$^{1,4}$, and
          D.~M.~Townsley$^{1}$} 
\vskip  0.15 true cm
\noindent{$^1$Supercomputer Computations Research Institute, 
          Florida State University, Tallahassee, Florida 32306-4052 }
\vskip -0.05 true cm
\noindent{$^2$Institute of Physics, Slovak Academy of Sciences,
          D\' ubravsk\' a cesta 9, 84228 Bratislava, Slovak Republic}
\vskip -0.05 true cm
\noindent{$^3$Department of Electrical Engineering, 
          2525 Pottsdamer Street,   
          Florida A\&M University--Florida State University,
          Tallahassee, Florida 32310-6046}
\vskip -0.05 true cm
\noindent{$^4$Center for Materials Research and Technology 
          and Department of Physics,
          Florida State University, Tallahassee, Florida 32306-3016}

\vskip 0.5 true cm

\noindent{\bf Abstract.}  
The magnetization switching dynamics in the kinetic Ising model
is projected onto a one-dimensional absorbing Markov chain.
The resulting projected dynamics reproduces the direct simulation
results with  great accuracy. A scheme is proposed to
utilize simulation data for small systems to obtain the metastable lifetime
for large systems and/or for very weak magnetic fields, for
which direct simulation is not feasible.

\vskip 0.2 true cm

%\noindent{\bf 1.~Introduction}
%\vskip 0.1 true cm
\noindent  
In simulations of metastable decay one faces the problem
of measuring the lifetime of the metastable phase, which is by definition
very long. For example, in Monte Carlo modeling of magnetization
switching in ferromagnets the physically relevant simulation
time scales are on the order of $10^{12}-10^{15}$ Monte Carlo Steps per Spin (MCSS).
Even with sophisticated algorithms [1,2]
such simulations have not yet been feasible, and one has to resort to
extrapolation of the results into the  physical
time-scale regime.

In this article, a scheme is presented to map the system under study
onto a simpler one, which is faster to simulate but still
gives accurate results for most important physical quantities. 
This idea is not new. For example, recently Lee et al. introduced
a macroscopic mean-field dynamics [3] which semiquantitatively describes
magnetization switching in Ising systems. The aim of the present work
is to construct a scheme which is very similar in spirit, but is
intended to create a practical computational tool applicable
to the simulation of ferromagnets in the physical time regime.

To simplify our notation and reasoning, we explain the
proposed Projected Dynamics (PD) as applied to the 
isotropic  kinetic Ising model on a square or cubic lattice.  
The Hamiltonian has the standard form,
$$
{\cal H} = -J \sum_{\langle ij \rangle} s_i s_j - H \sum_{i} s_i \ ,
\eqno(1) 
$$
with the ferromagnetic nearest-neighbor
spin-spin interaction $J>0$ and an external magnetic
field $H$. In what follows, we denote by $V$ the total number of spins in
the system. To study the magnetization reversal, we initialize all spins in
the state $+1$, fix the temperature $T$ well below its critical value $T_c$, and
apply a negative magnetic field. Then we apply Metropolis or Glauber
dynamics with updates at randomly chosen sites to measure the time the
system needs to reach a configuration with a given stopping  magnetization. 
Repeating
this procedure many times, we obtain the mean lifetime, $\tau$, of the
metastable state and its standard deviation, $\Delta\tau$. 

To speed up the simulations as much as possible, we use a rejection-free
algorithm [1,2] which uses the notion of spin classes.  By a spin class we mean
the state of the spin itself and its neighbors. There
are ten classes for the given model on a square lattice. 
Classes $i=1,\ldots,5$
correspond to spins in the state $+1$ which have exactly $i-1$ 
neighbors in the state $+1$.  Similarly classes
$i=6,\ldots,10$ are assigned to those $-1$ spins  which have $i-6$
neighbors in the state $+1$.  All spins in  class $i$ have the
same flipping probability, $p_i$. 

Rejection-free algorithms keep track of the number $c_i$ of spins in each
class, and it is not computationally expensive to measure the growth and
shrinkage rates of the stable phase, which are defined as 
$$ 
 g(n) = \sum_{i=1}^{5} \langle c_i \rangle_n p_i \ \ , \ \ 
 s(n) = \sum_{i=6}^{10}\langle c_i \rangle _n p_i \ ,
\eqno(2) 
$$ 
respectively. The angular brackets mean the average taken over the configurations
generated during the Monte Carlo lifetime experiment, conditionally
on the number $n$ of overturned spins.
Thus, $g(n)/V$ is the probability that a $+1$ spin
will be flipped in the next Monte Carlo step, and $s(n)/V$ is the
probability to flip one of the $-1$ spins, both conditionally on $n$. 
Flipping a $+1$ spin increases the 
volume fraction of the stable phase, hence the names $g(n)$ and $s(n)$.

The main idea of the proposed method is to make use of the observed 
growth and
shrinkage rates to map the switching dynamics onto a one-di- mensional
absorbing Markov chain. We assign to all configurations with $n$
overturned spins a single state $n$ in the chain. The one-dimensional
dynamics is given by the above probabilities: From the state $n$ we
have the probability $g(n)/V$ of jumping to the state $n+1$, the probability
$s(n)/V$ of jumping to $n-1$, and the probability $1-g(n)/V-s(n)/V$ of remaining in
the current state. This random walk starts at $n=0$ and terminates when it
reaches $n=N+1$ where $M=V-2N-2$ corresponds to the stopping magnetization. 
Using standard methods from the theory of absorbing Markov chains [2,3],
we obtain the mean
lifetime $\tau$ and the total average time $h(n)$ (in MCSS) spent by the
random walker in the state $n$ in terms of the growth and
shrinkage rates: 
$$ 
 \tau = \sum_{n=0}^N h(n) \ \ , \ \
 h(n) =  { 1 + s(n+1) h(n+1) \over  g(n)} \ \ , \ \ 
 h(N) =  {1\over g(N)} \ .
\eqno(3)
$$
Higher moments of the lifetime distribution can be obtained in a similar
way. 
\eject

\null
\vskip 6 truecm 
\includegraphics{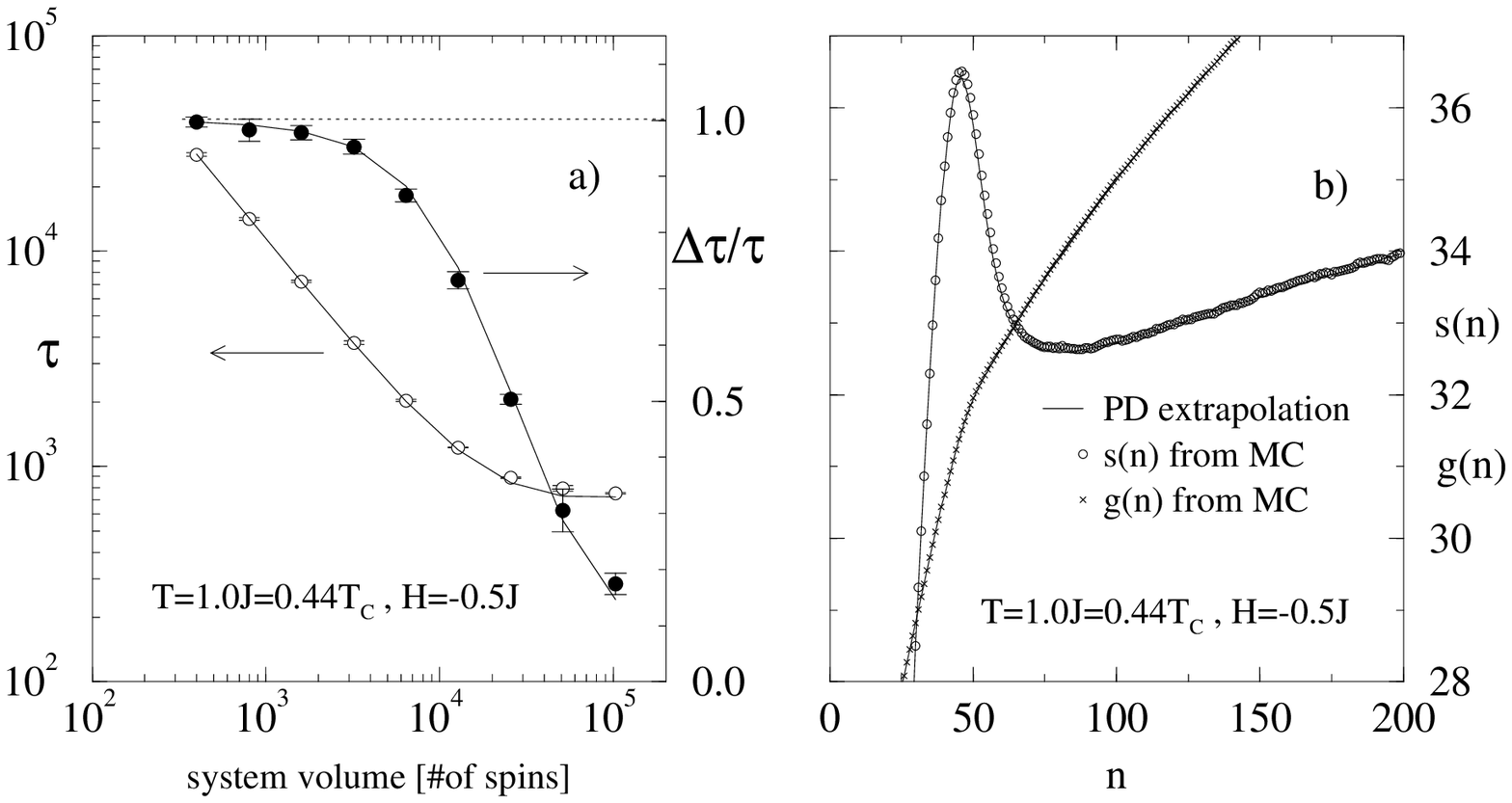} 

\noindent {\bf Fig. 1}~
Illustration of the Projected Dynamics [Eqs. (2,3)] and of the ``size extrapolation''
scheme [Eq. (4)]. The lines in both panels are Projected Dynamics extrapolations based
on $\langle c_i \rangle_n$ sampled during lifetime measurements on a $20\times20$ lattice.
Symbols are  direct simulation results.
a) The lifetime and its relative standard deviation vs. the system volume.
The agreement is near-perfect in the single-droplet regime (where $\Delta\tau/\tau\approx 1)$, 
and deviations are only observed in the crossover region to the multidroplet regime
(where $\Delta\tau/\tau\to 0$). 
b) Extrapolation from the $20\times20$ lattice reproduces the measured 
growth and shrinkage rates, $g(n)$ and $s(n)$, of the $160\times160$ lattice very well. 
The crossings of the two curves correspond to the metastable magnetization (left)
and the critical fluctuation (right).

\vskip 0.25 truecm

As shown in Fig.~1,
the proposed scheme reliably reproduces the direct Monte Carlo results. 
The lifetimes obtained from Eq.(3) are usually only slightly different from
their counterparts from the direct measurements.  The projected
dynamics also gives the standard deviation of the lifetime distribution,
which agrees with the measured values within the error bars.

Suppose we have measured $g(n)$ and $s(n)$ in a system of volume $V$. 
How are they related to their counterparts in a system of volume
$2 V$? Since the relevant configurations typically contain many small 
droplets of the stable phase, it is a reasonable approximation to view 
the larger system as consisting of two ``independent'' copies of the 
smaller system. Then, the growth rate of the large system can be 
approximated as a  weighted sum of the sub-system contributions,
$$
g(2V,n) \approx {\sum_{i=0}^n h(V,n-i) h(V,i) [ g(V,n-i) + g(V,i)] \over 
                 \sum_{i=0}^n h(V,n-i) h(V,i)} \ .
\eqno(4)
$$
Here, we have explicitly shown the dependence on the volume $V$. An analogous
formula is proposed for the shrinkage rate $s(n)$. 
Provided the smallest system is in the so-called single-droplet regime,
in which the nucleation is triggered 
{\parfillskip=0pt\par\parskip=0pt\noindent}

\null
\vskip 7 truecm 
\vskip -27 truept
\includegraphics{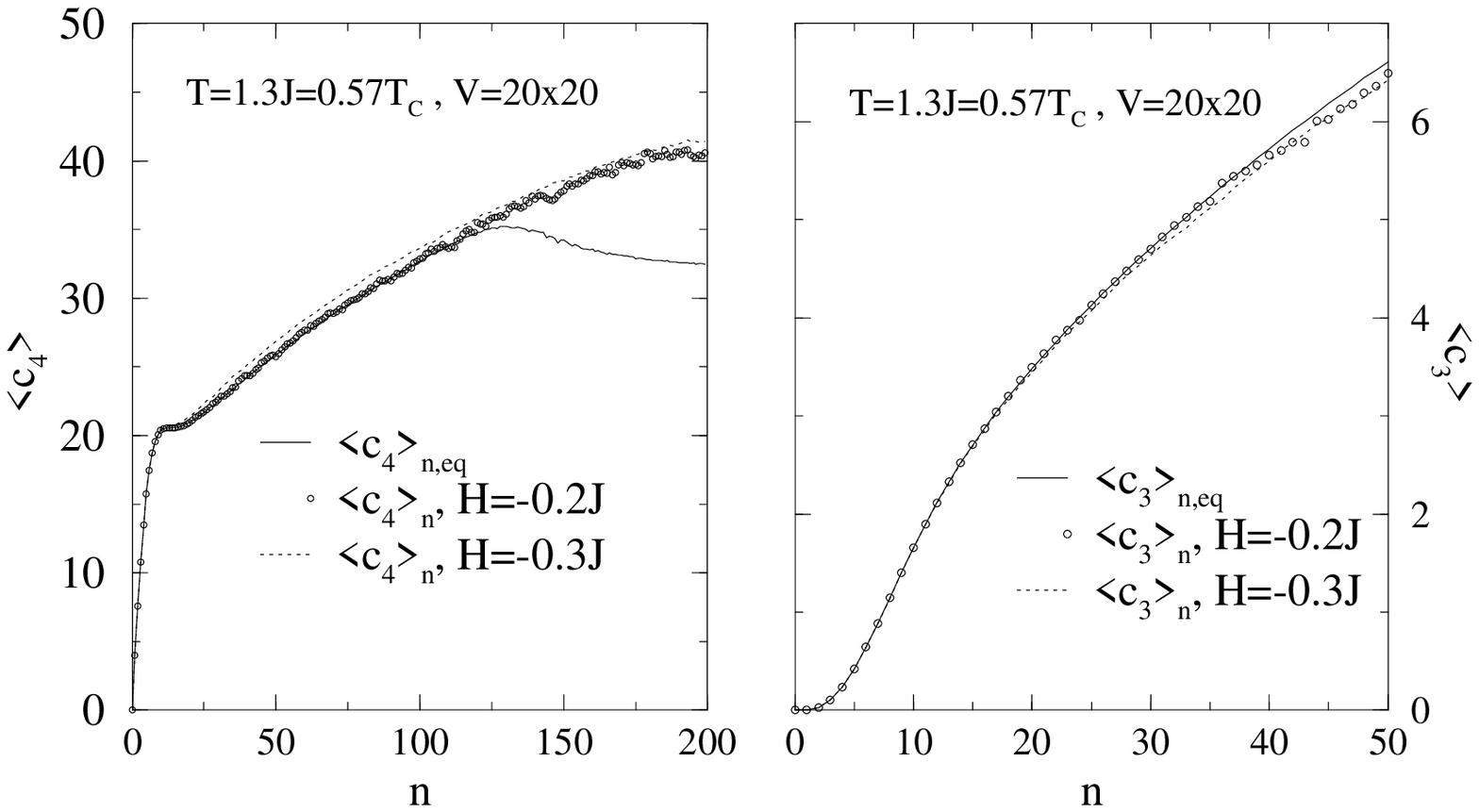} 

\noindent {\bf Fig. 2}~
Populations of two spin classes vs.\ the number of overturned spins $n$. Solid lines
are $\langle c_i \rangle_{n,{\rm eq}}$ measured in the equilibrium fixed-$n$ ensemble; circles and
dotted lines show $\langle c_i \rangle_{n}$ measured during the lifetime measurement for
two different fields. Note that as the field decreases the region in which 
$\langle c_i \rangle_{n} \to \langle c_i \rangle_{n,{\rm eq}}$ increases.

\vskip 0.25 truecm

\noindent
by a single critical fluctuation smaller than the system size
(for the theoretical description of different regimes of the magnetization
switching, see Refs. [4-8]), one
can repeat the extrapolation $V \to 2 V$ several times without finding a
big discrepancy between growth and shrinkage rates calculated from the
small-lattice data and those directly measured on large lattices
(see Fig. 1b).

Now we return to our main goal, namely predicting lifetimes in very
weak fields. What we need is to calculate $g(n)$ and $s(n)$ as  functions
of the magnetic field. The naive choice would be to replace the class
spin-flip probabilities $p_i$ by their values calculated for the desired
field, and keep the values $\langle c_i \rangle_{n}$ entering Eq. (2) unchanged.
Such a scheme works quite well when extrapolating to stronger fields, 
but the lifetimes are systematically
overestimated for fields weaker than the one at which the $\langle c_i \rangle_n$ were 
sampled.
The reason lies in the nonequilibrium character of the configurations with 
$n$ larger than the critical droplet volume (see Fig. 2). 
As it tries to escape from the metastable free-energy minimum,
when $n$ is small the system passes through configurations which 
are very close to ``equilibrium''. Farther from the free-energy minimum,
the configurations which appear in the system are increasingly different
from ``equilibrium'' configurations. Although these configurations 
do not contribute significantly 
to the lifetime, if we use them to estimate the lifetime in a
{\it weaker} field their contribution becomes more important because the system
then spends more time in them. However, the actual configurations are closer
to ``equilibrium,'' because the free-energy minimum is deeper in a weaker field.  

\eject
\null
\vskip 6.25 truecm
\vskip -22 truept 
\includegraphics{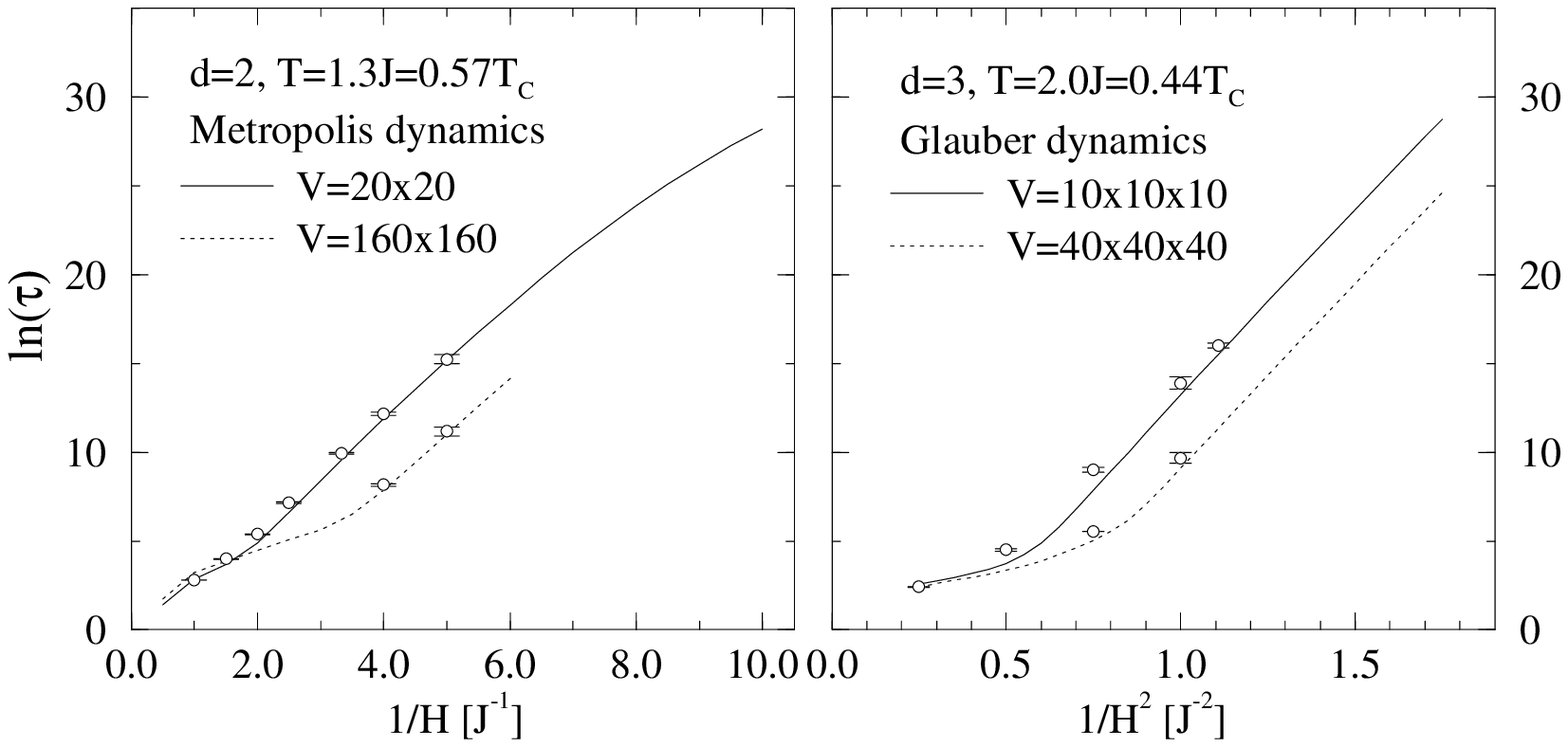} 

\noindent {\bf Fig. 3}~
The metastable lifetime of a kinetic Ising model as a function of the magnetic field.
Points are conventional Monte Carlo measurements, and lines are Projected Dynamics calculations
based on the ``equilibrium'' $\langle c_i \rangle_{n,{\rm eq}}$ data sampled on the smaller systems. 
Predictions of the Projected Dynamics improve with decreasing field.

\vskip 0.25 truecm

\noindent
Thus, we effectively replace the actual weak-field
configurations by ones which appear slightly 
``overheated.'' 
This enhances the shrinkage probabilities more than the
growth probabilities, thus leading to an overestimate of the
metastable lifetime. 
Understanding the cause of this problem also offers a remedy: we
can do better by using the equilibrium configurations. Then we
expect to see underestimation of relatively short lifetimes (by reversing
the above argument). On the other hand, such an approximation will improve with
decreasing field, which is exactly what we need to extrapolate towards zero
field. Thus, we define the equilibrium growth and shrinkage
rates as follows
$$
g_{\rm eq}(n) = \sum_{i=1}^{5} \langle c_i \rangle_{n,{\rm eq}} p_i 
\ \ , \ \
s_{\rm eq}(n) = \sum_{i=6}^{10}\langle c_i \rangle_{n,{\rm eq}} p_i \ .
\eqno(5)
$$
The only difference from Eq.~(2) is in the sampling. Here, $\langle c_i \rangle_{n,{\rm eq}}$
are sampled in an equilibrium ensemble with a fixed number of overturned spins  $n$.
Thus, instead of sampling the configurations during the lifetime measurement, we
perform a static measurement for each value of $n$ needed (up to the stopping magnetization).
The $\langle c_i \rangle_{n,{\rm eq}}$ depend on temperature, but the only dependence of 
$g_{\rm eq}$ and $s_{\rm eq}$ on the
magnetic field comes from the spin-flip probabilities $p_i$. In that way, a single measurement provides 
sufficient data  to obtain a good approximation for
the lifetime as a function of the external field.

Figure 3 shows the lifetime calculated from the projected dynamics for two pairs of lattice sizes
in two and three dimensions.
Comparison with the direct simulation results (points) corroborate our expectation that the
lifetime is underestimated in the crossover region between the single-droplet and
multidroplet regimes, and that the approximation provides progressively better results
as the external field strength decreases. Measurements on small lattices, for which we can
obtain reasonable statistics in zero field, show that the Projected Dynamics based on
$\langle c_i \rangle_{n,{\rm eq}}$ reproduces the ``lifetime'' in zero field.

In conclusion, our Projected Dynamics maps the complex Monte Carlo dynamics onto
a much simpler one-dimensional absorbing Markov chain. 
The Projected Dynamics is computationally much easier to study than the full underlying
Monte Carlo dynamics, while
it provides reliable results for metastable lifetimes and their standard deviations. 
Data from small systems can be utilized to predict lifetimes for
large systems. When based on ``equilibrium'' class populations 
$\langle c_i \rangle_{n,{\rm eq}}$, Projected Dynamics yields results
as functions of the field. Such estimates are most reliable in the experimentally 
relevant weak-field region, which it is currently not feasible to investigate with 
direct simulations. 
The method also offers a deeper insight into the mechanism of metastable decay.
For example, the histogram $h(n)$ provides a simple method to measure
the metastable magnetization and susceptibility, and the rates $g(n)$ and $s(n)$ 
contain information about the size of the critical fluctuations.

For other types of models, the applicability of the projected dynamics
will depend on the existence of a ``single channel'' for the decay. The projected parameter
($n$ in the present case) need not necessarily be related to the magnetization, but it should
parameterize the optimal path for escape from metastability in such a way that the
configurations at a fixed value of the parameter will exhibit similar growth and shrinkage
rates.

We dedicate this paper to Professor Masuo Suzuki on his 60th birthday. This research was
supported by FSU-MARTECH, FSU-SCRI (DOE Contract No. DE-FC05-85ER25000),
NSF Grants No. DMR-9315969, DMR-9520325, and DMR-9634873.
\vskip 0.15 truecm

\noindent{\bf References}
\vskip 0.1 truecm

\noindent [1]~A.\ B. Bortz, M.\ H. Kalos and J.\ L. Lebowitz, J. Comput. Phys. {\bf 17}, 10 (1975).

%\bibitem{Novotny}
\noindent [2]~M.\ A. Novotny, Phys.\ Rev.\ Lett. {\bf 74}, 1 (1995);
Erratum {\bf 75}, 1424 (1995).

\noindent [3]~J.\ Lee, M.\ A. Novotny and P.\ A. Rikvold, Phys.\ Rev.\ E {\bf 52}, 356 (1995).
%\bibitem{Rikvold}

\noindent [4]~P.~A.\ Rikvold, H.\ Tomita, S.\ Miyashita and S.~W.\ Sides,
Phys.\ Rev.\ E {\bf  49}, 5080 (1994).

%\bibitem{Richards95}
\noindent [5]~H.~L.\ Richards, S.~W.\ Sides, M.~A.\ Novotny, and P.~A.\ Rikvold,
J.\ Magn.\ Magn.\  Mater. {\bf 150}, 37 (1995). 

%\bibitem{Richards96a}
\noindent [6]~H.~L.\ Richards, S.~W.\ Sides, M.~A.\ Novotny, and P.~A.\ Rikvold,
J.\ Appl.\ Phys.\ {\bf 79}, 5479 (1996).

%\bibitem{Richards96b}
\noindent [7]~H.~L.\ Richards, M.~A.\ Novotny, and P.~A.\ Rikvold,
Phys.\ Rev.\ B {\bf  54}, 4113 (1996).

%\bibitem{Richards97}
\noindent [8]~H.~L.\ Richards, M.\ Kolesik, P.-A.\ Lindg\aa rd, P.~A.\ Rikvold, and M.~A.\ Novotny,
Phys.\ Rev.\ B, {\bf 55} in press.

\end